\documentclass{iucr}              % DO NOT DELETE THIS LINE

\usepackage{graphicx}% Include figure files
\usepackage{bm}
\usepackage{float}
\usepackage[usenames]{color}
\graphicspath{{./figures/}}

     \journalcode{J}              % Indicate the journal to which submitted
\begin{document}                  % DO NOT DELETE THIS LINE

     %-------------------------------------------------------------------------
     % The introductory (header) part of the paper
     %-------------------------------------------------------------------------

     % The title of the paper. Use \shorttitle to indicate an abbreviated title
     % for use in running heads (you will need to uncomment it).

\title{xPDFsuite: an end-to-end software solution for high throughput pair distribution function transformation, visualization and analysis}
\shorttitle{xPDFsuite}

     % Authors' names and addresses. Use \cauthor for the main (contact) author.
     % Use \author for all other authors. Use \aff for authors' affiliations.
     % Use lower-case letters in square brackets to link authors to their
     % affiliations; if there is only one affiliation address, remove the [a].

\author[a]{Xiaohao}{Yang}
\author[b]{Pavol}{Juh\'{a}s}
\author[a]{Christopher~L.}{Farrow}
\cauthor[a,b]{Simon~J.~L.}{Billinge}{sb2896@columbia.edu}

\aff[a]{{Department of Applied Physics and Applied Mathematics,
    Columbia University},
    \city{{New York}, New York, 10027, \country{USA}}}
\aff[b]{{Condensed Matter Physics and Materials Science Department,
    Brookhaven National Laboratory},
    \city{Upton, New York, 11973, \country{USA}}}
     % Use \shortauthor to indicate an abbreviated author list for use in
     % running heads (you will need to uncomment it).

%\shortauthor{Soape, Author and Doe}

     % Use \vita if required to give biographical details (for authors of
     % invited review papers only). Uncomment it.

%\vita{Author's biography}

     % Keywords (required for Journal of Synchrotron Radiation only)
     % Use the \keyword macro for each word or phrase, e.g.
     % \keyword{X-ray diffraction}\keyword{muscle}

%\keyword{keyword}

     % PDB and NDB reference codes for structures referenced in the article and
     % deposited with the Protein Data Bank and Nucleic Acids Database (Acta
     % Crystallographica Section D). Repeat for each separate structure e.g
     % \PDBref[dethiobiotin synthetase]{1byi} \NDBref[d(G$_4$CGC$_4$)]{ad0002}

%\PDBref[optional name]{refcode}
%\NDBref[optional name]{refcode}

\maketitle                        % DO NOT DELETE THIS LINE

\begin{synopsis}
xPDFsuite is a software program for processing and analyzing pair distribution function data, including 2D image processing and integration, pair distribution function transformation, and atomic structure modeling. 
\end{synopsis}

\begin{abstract}

The xPDFsuite software program is described. It is for processing and analyzing atomic pair distribution functions (PDF) from X-ray powder diffraction data. It provides a convenient GUI for SrXplanr and PDFgetX3, allowing the users to easily obtain 1D diffraction pattern from raw 2D diffraction images and then transform them to PDFs. It also bundles PDFgui which allows the users to create structure models and fit to the experiment data. It is specially useful for working with large numbers of datasets such as from high throughout measurements. Some of the key features are: real time PDF transformation and plotting; 2D waterfall, false color heatmap, and 3D contour  plotting for multiple datasets; static and dynamic mask editing; geometric calibration of powder diffraction image; configurations and project saving and loading; Pearson correlation analysis on selected datasets; written in Python and support multiple platforms.

\end{abstract}

     %-------------------------------------------------------------------------
     % The main body of the paper
     %-------------------------------------------------------------------------
     % Now enter the text of the document in multiple \section's, \subsection's
     % and \subsubsection's as required.

\section{Introduction}
\label{Introduction}

Increasingly, scientists are interested in materials that have local structure at the nano-scale~\cite{billi;p10}. However it is difficult to do quantitative studies of the nano-scale structure using conventional diffraction techniques since the Bragg diffraction peaks are extremely broad, or absent.~\cite{billi;s07} To address this problem, atomic pair distribution function (PDF) analysis is becoming popular~\cite{young;jmc11,billi;jssc08} which uses both Bragg and diffuse scattering, and Fourier transforms the diffraction data to real space allowing local structure to be studied, even from nanoparticles and molecules.~\cite{egami;b;utbp12}

To obtain the PDF from experimental data, the diffraction Intensity $I(Q)$ is first properly corrected and normalized to the total scattering function $S(Q)$, and then Fourier transformed to PDF using,
\begin{equation}
\centering \label{equ;sqtogr}
G\left(r\right) = {2\over\pi}\int_{Q_{min}}^{Q_{max}} Q[S(Q)-1] \sin Qr \>  \mathrm{d}Q,
\end{equation}
where $Q=4\pi\sin\theta/\lambda$ is thte magnitude of the scattering vector (here $\theta$ is half the scattering angle and $\lambda$ is the x-ray wavelength). $Q_{min}$ and $Q_{max}$ are the limits of the $Q$ range used in the transformation. We also define the reduced scattering function $F(Q)$ as $F(Q) = Q[S(Q)-1]$. Many corrections must be carried out to obtain these functions~\cite{egami;b;utbp12}.  However, recently an ad hoc approach for carrying out these corrections was described~\cite{billi;jpcm13}, which has been implemented in a software program, PDFgetX3~\cite{juhas;jac13}.  This new approach is rapid, making feasible high throughput PDF analyses, such as {\it in-situ} studies with many data-points~\cite{jense;jacs12,tyrst;acie12} and even imaging using computed tomography PDFs~\cite{jacqu;nc13}. 

A full PDF data analysis usually starts from the 2D powder diffraction images, which later been azimuthally integrated into 1D diffraction pattern and then transformed to PDFs using Fourier transformation. An atomic structure model could be build and fitted to the data to obtain the quantitative structure information. This paper describe an end-to-end software solution, xPDFsuite, that provides all these functionalities in a easy to use graphical user interface (GUI). xPDFsuite warps SrXplanar~\cite{yang;jac14} to process 2D diffraction image, PDFgetX3~\cite{juhas;jac13} to transform PDF, and PDFgui~\cite{farro;jpcm07} to build and fit the structure model. It not only implemets all functionality of these software, but it also provides many extra functions help the user to obtain and visualzie high quality PDFs easily and quickly from many samples. 

\section{User interface and Capabilities}

\subsection{Overview of user interface}

\begin{figure}
	\caption{a) SrXgui main window, and its image and mask editing panel. b) xPDFsuite main window, and plot window of single example data set, including $I(Q)$, and $G(r)$. c) PDFgui and plot of one refinement d) Other utilities, including Pearson correlation analysis and inter-atomic distance calculator. }
	\label{fig;maingui}
	\includegraphics[width=1.0\textwidth, keepaspectratio]{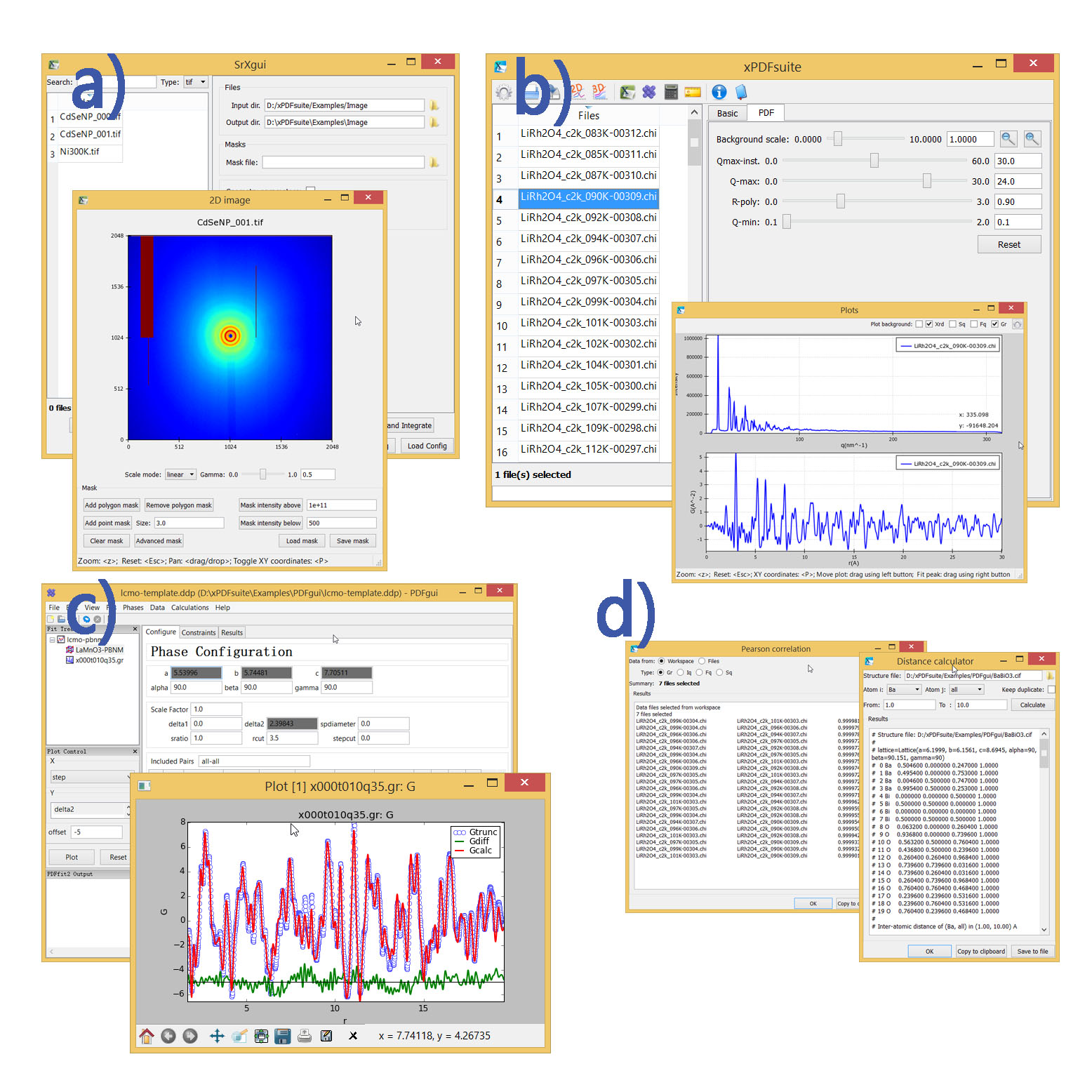}
\end{figure}

The xPDFsuite has four major components, as shown in Fig.~\ref{fig;maingui}: Main window, that allows the user to start all other components and to perform PDF transformations; SrXgui, which processes 2D diffraction image and produces 1D diffraction pattern using SrXplanar; PDFgui, that allows users to do quantitative structure modeling for PDF data; and other utilities, including Pearson correlation analysis and inter-atomic distance calculator. In main window, all other components and most frequently used functions can be started by clicking the corresponding buttons in the toolbar or in the dropdown menu. 

\subsection{2D diffraction image processing: SrXgui}

SrXgui uses SrXplanar as integration engine to process 2D powder diffraction image. The 2D images can be azimuthally integrated into 1D pattern using non-splitting pixel algorithm~\cite{yang;jac14}. The geometric parameters of 2D detector should be provided directly or calibrated using a 2D image. SrXgui supports two calibrating modes: self-mode and calibrant-mode. Self-mode calibrates the image by varying the beam center until it can get the sharpest peak, therefore it doesn’t require a specific calibrant. If diffraction image from a standard calibrant is available, then user can use calibrant-mode to obtain the correct geometric parameters. The software uses pyFAI~\cite{kieff;jpconfs13} as calibration engine. 

SrXgui can create and edit mask to mask bad pixels, such as shadow of sample holder and dead pixels. In SrXplanar, there are two types of masks, static mask and dynamic mask. Static mask is mask that doesn't change during integration, and it is usually used to mask sample holder shadow and dead pixels. In contrast, dynamic mask is generated for each image during integration. It can mask the pixels that are much darker or much brighter compared to their neighbors or the average intensity for all pixels at similar diffraction angle. The dynamic mask is useful for masking the spots in spotty data. 

Once the geometric parameters and mask are specified, user can select and integrate image files, either integrate separately, i.e. each image produces one diffraction pattern, or sum and integrate, i.e. all selected images are summed together and then produce only one pattern. The integrated file will be automatically imported to the main xPDFsuite GUI for the following data reduction.

\subsection{PDF transformation: main GUI}

xPDFsuite main window is used to control analysis and to spawn one or many plot windows, as shown in~\ref{fig;maingui} b). The main window has a workspace panel which allows the user to select data sets to be processed,  and a control panel that allows the user to change the parameters of the PDF transformations. The different types of data in single or multiple data sets can be plotted in plot panels simultaneously and parameters updated through the use of sliders or by typing values into a field. 

\subsubsection{Real time transformation and plots}

The design of xPDFsuite allows the user to see the results of their actions in real time, i.e. the plots refresh automatically after each user input. This is useful when exploring the parameter set for PDF transformations, including but not limited to the scale-factor of a background signal being subtracted, $Q_{max-inst}$,$Q_{max}$, $Q_{min}$, $Rploy$. Please see the documentation of PDFgetx3~\cite{juhas;jac13} and xPDFsuite for the meaning of these parameters.

\subsubsection{Data sets selection}

The 1D diffraction data sets can either be integrated using SrXgui from 2D images or imported to xPDFsuite directly. In later case, the data files to be processed should be in the form of multi-column ascii files. They can be added to the workspace through "add files" pop-up window. Filter functions, incorporating simple regular expressions, are available for finding desired files in directories containing large numbers of files.  Processed data can also be added to workspace but only for plotting. 

The data sets added to the workspace can be selected and plotted in new plot panels. Only selected data sets or data set plotted in the plot panel will be recalculated when transformation parameters are changed, to accelerate the data processing and plotting. By default, the data set(s) plotted in one existing plot are not changed after the plot is initialized; however a "sync" mode is available in the drop down menu in which the data set(s) are synced with the selection in workspace.

\subsubsection{Plot panels}

xPDFsuite provides flexible ways of plotting data sets. For each single data set, there are four types of data can be plotted in xPDFsuite, including raw X-ray intensity $I(Q)$, total scattering function $S(Q)$, reduced scattering function $F(Q)$, and final PDF $G(r)$. User can turn on or off any type of plot in the plot panel. Again, like the adjusting the PDF transformation parameters, the plots are updated on the fly after any changes have been made. All type of data can be exported as pure txt-file and the plots can be exported as image files using save functions in the toolbar of the main window, or in the dropdown menu of plot panel.

If only a single data set is selected and plotted in a plot panel, and the X-ray intensity $I(Q)$ is selected, the user can additionally overlay the scaled background signal on the $I(Q)$ plot, which is also updated on the fly when the background scale is changed. This is helpful when adjusting the background scale factor.

xPDFsuite also supports plotting multiple data sets simultaneously in one plot panel, such as plotting a series of data sets collected under different temperature. There are two modes of multiple data set plotting: 2D and 3D, which may are chosen when the plot panel is created.
\begin{figure}
\caption{The xPDFsuite plot panels in 2D and 3D mode. The example data sets are PDF of LiRh$_2$O$_4$ in different temperature ranging from 80K to 300K with 2K interval. a) 2D plot of 7 sets, b) 2D waterfall plot of 111 data sets, c) false color heat map of 111 data sets, d) 3D contour plot}
\label{fig;plot}
\includegraphics[width=1.0\textwidth, keepaspectratio]{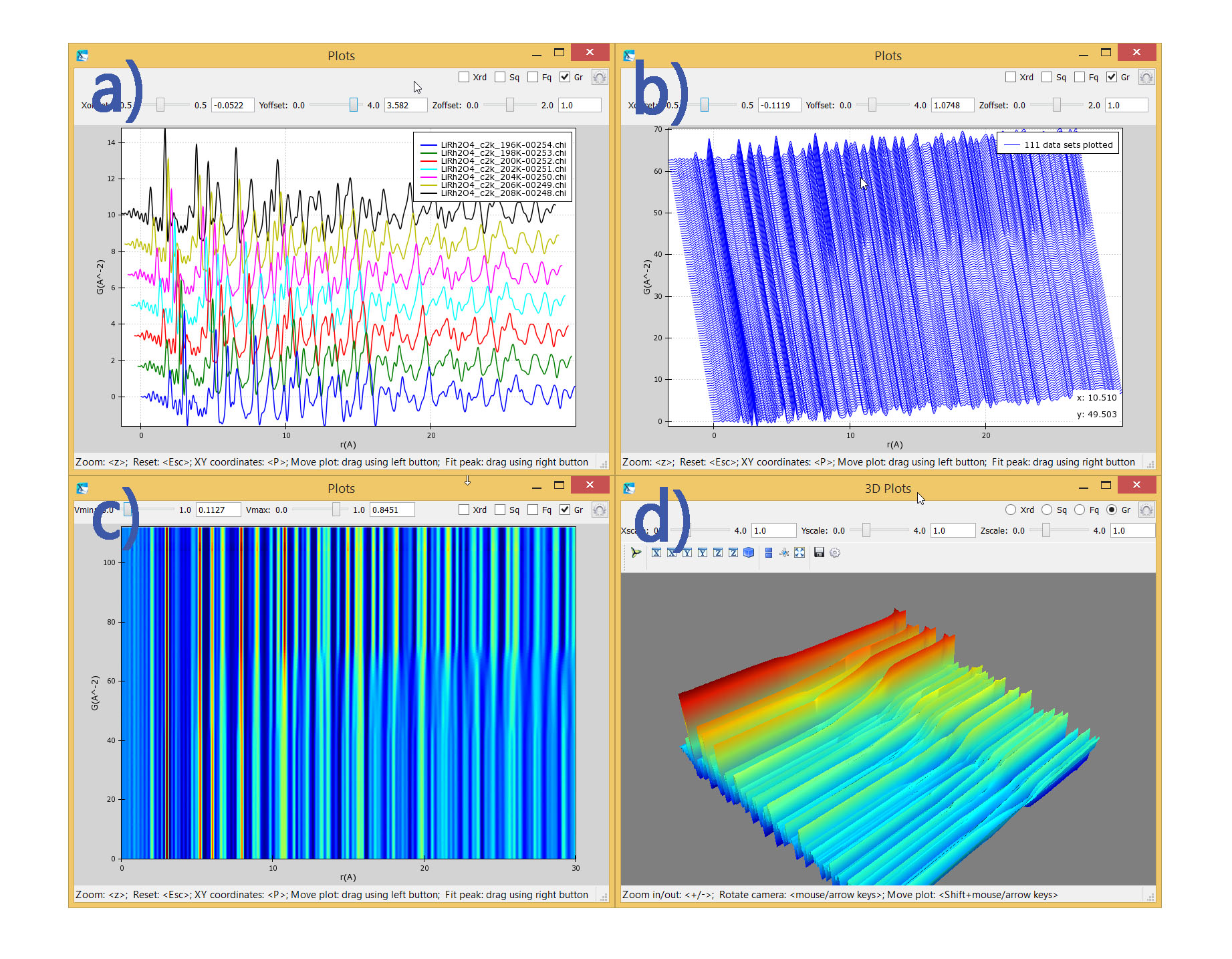}
\end{figure}

In 2D mode, all data sets could be overlaid in a water-fall style, as shown in Fig.~\ref{fig;plot} a) and b). The user can adjust the distance between curves, and the tilt of the false 3D view. If the number of data sets is below 8, different data sets will be colored and a color legend will be displayed. When the number exceeds 8, to improve the performance, all data sets are colored blue. The user can manually switched between multiple color mode (slower) and single color mode (faster) by toggle 'fastmode' in the dropdown menu.

Alternatively, user can also have a "heatmap" generated from multiple data sets, as shown in Fig.~\ref{fig;plot} c). The range of data to be color mapped could be changed. Any data points excess the range will be mapped to red (greater than upper bound) or blue (smaller than lower bound). 

In 3D mode, Fig.~\ref{fig;plot} d), the selected data-sets are stacked into a 2D array and then a 3D surface is rendered using this 2D array. The user can rotate, pan, or zoom the 3D surface freely, as well as adjusting the scale of the 3D surface in its 3 directions. Since the number of data points, i.e. number of vertices in the surface, is very large when a large number of data sets are selected, the software under-samples the surface to accelerate the 3D rendering, which may result in a rough surface. The user can disable this behavior by turning off 'fastmode' in the dropdown menu.

\subsection{Structure modeling: PDFgui}

To analysis and model the PDF data, user could use PDFgui~\cite{farro;jpcm07}, which provides convenient GUI for building atomic structural model and fitting the models to experiment data. The PDFgui is bundled in xPDFsuite and could be called from main GUI of xPDFsuite.

\subsection{Tools and utilities}

\subsubsection{Pearson correlation analysis}

Pearson correlation function is quite frequently used in comparing data sets. %It is calculated as, 
%
%\begin{equation}
%P = \frac{\sum(x-\bar{x})(y-\bar{y})}{\sqrt{\sum(x-\bar{x})^2}\sqrt{\sum(y-\bar{y})^2}}, 
%\end{equation}
%
%where $x$ and $y$ are two data sets and the summation goes over all the data points. 
In xPDFsuite, user can choose data sets in the workspace, in which case all four types of data are available to compare, or choose a directory, where all data files to be compared located. Program will calculate the Pearson correlation coefficient of all possible combinations and sorted by the correlation coefficient. The results can be copied to clipboard or exported as pure txt file. 

\subsubsection{Inter-atomic distance calculator}

The user can calculate the inter-atomic distance for a given structure using this tool. The calculation range and the type of elements in atom pairs can be specified. Similar to the Pearson correlation analysis, the results can be either copied to clipboard or exported as pure txt file.

\section{Documentation, availability and environment}

A graphic tutorial and a graphic reference manual are included in the main GUI and the example data sets (also used in this paper) are distributed with the software. 

xPDFsuite is written in Python language. It runs on Windows, Linux, MacOSX, and all major Unix system. We provide all-in-one 64bit pre-build bundle for the Windows, Linux, and MacOSX system. 

xPDFsuite requires a license from Columbia University. Please contact Prof. Billinge (sb2896@columbia.edu) or consult the web page (www.diffpy.org/products/xPDFsuite). PDFgetX3-academic, a command line version of PDFgetX3 may be downloaded for free and used, subject to the license conditions, by academics and researchers at non-profit institutions for open research.  All other uses require a commercial license for PDFgetX3 available from Columbia University. SrXplanar (avaiable at github.com/diffpy/diffpy.srxplanar) and PDFgui (available at www.diffpy.org) are open-soure software developed by Billinge group. More information is available from Prof. Billinge at sb2896@columbia.edu.

     % Appendices appear after the main body of the text. They are prefixed by
     % a single \appendix declaration, and are then structured just like the
     % body text.

     %-------------------------------------------------------------------------
     % The back matter of the paper - acknowledgements and references
     %-------------------------------------------------------------------------

     % Acknowledgements come after the appendices

\ack{Acknowledgements}

We would like to thank Maxwell Terban all other Billinge group members for help and useful suggestion during the development and testing. This work is supported as a Laboratory Directed Research and Development (LDRD) Program 12-007 (Complex Modeling) at Brookhaven National Laboratory, which is funded by the US Department of Energy Office of Basic Energy Sciences grant DE-AC02-98CH10886.

     % References are at the end of the document, between \begin{references}
     % and \end{references} tags. Each reference is in a \reference entry.

     %-------------------------------------------------------------------------
     % TABLES AND FIGURES SHOULD BE INSERTED AFTER THE MAIN BODY OF THE TEXT
     %-------------------------------------------------------------------------

     % Simple tables should use the tabular environment according to this
     % model

     % Postscript figures can be included with multiple figure blocks

\bibliographystyle{iucr}
%\bibliography{billinge-group,abb-billinge-group,everyone}
\bibliography{getxgui}

\end{document}